\documentclass[twocolumn,showpacs,preprintnumbers,amsmath,amssymb]{revtex4}
\usepackage{dcolumn}
\usepackage{bm}
\usepackage{graphicx}
\usepackage{amsmath}
\begin{document}

\title{Exact coherent matter-wave solitons induced and controlled by laser field}
\author{Wenhua Hai$^*$, \ \ Qiongtao Xie, \ \ Qianquan Zhu}
\affiliation{Key Laboratory of Low-dimensional Quantum Structures
and Quantum Control of Ministry of Education, and
\\ Department of Physics, Hunan Normal University, Changsha 410081, China} 
\email{whhai2005@yahoo.com.cn}

\begin{abstract}

We find a set of exact solutions of coherent bright solitons in the
quasi-one-dimensional (1D) Bose-Einstein condensate (BEC) trapped in
a harmonic potential, by using a Gaussian laser well (barrier) with
oscillating position to balance the repulsive (attractive)
interatomic interaction. The bright solitons do not deform in
propagation and are controlled accurately by the laser driving which
resonates with the trapping potential. The solitonic motion is more
stable for the repulsive BEC than that of the attractive BEC. The
results reveal a different kind of soliton trains compared to that
reported recently in Phys. Rev. Lett. 100, 164102 (2008) and suggest
an experimental scheme for generating and controlling the coherent
matter-wave solitons.

\end{abstract}


\maketitle

Soliton in a Bose-Einstein condensate (BEC) is a kind of important
nonlinear phenomena, which has been investigated experimentally and
theoretically \cite{Strecker}-\cite{Huang}. The formation and
propagation of BEC solitons were observed by magnetically tuning the
atom-atom interaction from repulsive to attractive \cite{Strecker,
Cornish}. The theoretical works demonstrated that the matter-wave
bright solitons can be created in a BEC, through the modulational
instability \cite{Carr, Salasnich} and quantum phase fluctuations
\cite{Khawaja}. Propagation feature of the solitons is the breathing
oscillation which is mostly controlled by the harmonic trap
\cite{Adhikari, Hai}. In the harmonic trapping case, the Gaussian
ansatz was used to fit the profiles of solitons \cite{Khaykovich,
Khawaja, Pedri, Dalfovo}. The Gaussian-shaped optical potentials
have been applied for investigating the BEC solitons \cite{Beitia,
Rodas, Lee} and quantum tunneling \cite{Zollner,Salasnich2, Coullet,
Luo}. It is worth noting that the balance between nonlinearity and
dispersion was found in the seminal study of soliton \cite{Zabusky}.
Recently, the new balances between the atom-atom interaction and the
Gaussian and/or periodical potentials are demonstrated \cite{Rodas,
Hai2}. For some special forms of external potential and interaction
intensity, the exact soliton solutions in BECs have also been
reported \cite{Beitia, Beitia2, Liang, Chong}.

The quantum states governed by the linear Schr\"{o}dinger equation
with inseparable space-time variables are very important but had to
find. The coherent state of a harmonic oscillator is a nice example
of such states, which has been widely applied to physics and optics
\cite{Glauber,Klauder} and is also extended to the case of
wavepacket trains \cite{Hai,Hai3}. Can the coherent wavepacket
trains exist in a harmonically trapped BEC system governed by the
Gross-Pitaevskii equation (GPE)? This is usually impossible, because
of the nonlinearity in GPE. However, when we employ the laser field
to balance the nonlinear term, seeking exact coherent states of the
GPE could become possible. Demonstrating the exact coherent state of
GPE and its experimental feasibility is our main motivation in this
paper.

By using the balance technique and applying the oscillating Gaussian
lasers, we find $n$ exact solutions of coherent soliton trains in
the quasi-1D BEC. When $n=0$ is considered, for the attractive or
repulsive BEC the interatomic interaction is balanced by the
Gaussian barrier or well respectively, and in the both cases the
soliton solutions possesses the same form of coherent state. The
coherent bright soliton oscillates like a classical harmonic
oscillator with the trapping frequency, which agrees with Strecker's
experimental results \cite{Strecker}. However, compared to the
deformed soliton trains observed in experiment, our bright solitons
have different properties, namely their shapes are kept in
propagation, their behaviors are controlled accurately by the laser
field and their motions possess more stability for the repulsive BEC
rather than the attractive one. Based on the capacity of current
experiments, such bright solitons can be observed in a BEC.

We consider a BEC consisting of $N$ identical Bose atoms and being
transferred into a cigar-shaped harmonic trap. The potential that
takes into account the combination of the magnetic trap with the
laser sheet reads $V(x,t)=\frac {1}{2}m \omega_x^2 x^2+V_L(x,t)$
with $V_L(x,t)$ containing the Gaussian-shaped laser potential
\cite{Beitia, Coullet, Luo}. Let the transverse frequency $\omega_r$
be much greater than the axial frequencies $\omega_x$, the dynamics
of the system is governed by the quasi-1D GPE \cite{Khawaja,
Dalfovo}
\begin{equation}
i \hbar \frac{\partial \psi}{\partial t}=- \frac {\hbar^2}{2m}
\frac{\partial ^2\psi}{\partial x ^2} + [V(x,t)
+g'_{1d}|\psi|^2]\psi,
\end{equation}
where we have assumed the transverse wave function being in ground
state of a harmonic oscillator such that the quasi-1D interaction
intensity related to the $s$-wave scattering length $a_s$, atomic
mass $m$ and number of condensed atoms $N$ reads \cite{Gardiner}
$g'_{1d}=Nm \omega_r g_0/(2\pi \hbar)= 2 N \hbar \omega_r a_s$ for
the normalized wave-function $\psi$. The norm $|\psi|^2$ is the
probability density and $N|\psi|^2$ the density of atomic number.
Setting $l_r=\sqrt{\hbar /(m \omega_r)}, \ l_x=\sqrt{\hbar /(m
\omega_x)}$, we normalize the time, space, wave function and laser
potential by $\omega_x^{-1},\ l_x,\ \sqrt{1/l_x}$ and $\hbar
\omega_x$ respectively, then the interaction intensity becomes
$g_{1d}=2 N \omega_r a_s/(\omega_x l_x)$ and the dimensionless GPE
reads
\begin{eqnarray}
i \frac{\partial \psi}{\partial t}= - \frac {1}{2} \frac{\partial
^2\psi}{\partial x ^2} + \Big[\frac {1}{2} x^2+V_L(x,t)
+g_{1d}|\psi|^2\Big]\psi.
\end{eqnarray}
In order to seek the extended coherent state of Eq. (2), we have to
use the balance condition \cite{Hai2}
\begin{equation}
V_L(x,t)+g_{1d}|\psi|^2 =\mu
\end{equation}
to transfer Eq. (2) to the linear Schr\"{o}dinger equation
\begin{eqnarray}
i \frac{\partial \psi}{\partial t}=- \frac {1}{2} \frac{\partial
^2\psi}{\partial x ^2} + \Big[\frac {1}{2} x^2+\mu \Big]\psi,
\end{eqnarray}
where $\mu$ denotes an undetermined constant determined by the laser
profile and is in units of $\hbar \omega_x$. Obviously, by the
balance condition we mean that the external optical potential and
the internal interaction reach into an indifferent equilibrium
experimentally, namely their sum equates a constant \cite{Hai2}. The
exact solutions of Eq. (2) must obey the balance condition (3) and
linear equation (4) simultaneously. Therefore, only the properties
common to the nonlinear Eq. (3) and linear Eq. (4) can be kept in
the balance solution. For the Gaussian potential $V_L(x,t)$ the
soliton solution of the nonlinear Eq. (3) and coherent-state
solution of the linear Eq. (4) can be in the same form and could
coexist thereby.

The exact solutions of extended coherent states of Eq. (4) read
\cite{Hai,Hai3}
\begin{eqnarray}
\psi_n&=&R_n(x,t)e^{i\Theta_n(x,t)},\ \ \ n=0,1,2,\cdots; \nonumber \\
R_n&=&\Big(\frac{1}{\sqrt{\pi}2^nn!}\Big)^{1/2}H_n(x-x_0\cos t)e^{-\frac 1 2(x-x_0\cos t)^2},\nonumber \\
\Theta_n&=&-\Big[\Big(\frac 1 2+\mu+n\Big)t+x_0 x \sin t-\frac 1 4
x_0^2\sin 2t\Big],
\end{eqnarray}
which can be proved by inserting it directly into Eq. (4). Here
$H_n$ denotes the Hermite polynomial with $x_0$ being the amplitude
of the center position of Gaussian packet, which can be adjusted by
the oscillating amplitude of laser position. Applying Eq. (5) to Eq.
(3) shows the profile of required laser field
\begin{equation}
V_L =\mu-\frac{g_{1d}}{\sqrt{\pi}2^nn!}H_n^2(x-x_0\cos
t)e^{-(x-x_0\cos t)^2}.
\end{equation}
Clearly, for constant interaction $g_{1d}$ and $n=0$ $(H_0=1)$ case
Eq. (6) describes the oscillating Gaussian potential which agrees
with that used in \cite{Beitia, Smerzi}. At $x_0=0$ it becomes the
well known time-independent form \cite{Zollner, Rodas, Coullet}
which leads Eq. (5) to the exact stationary state of Eq. (2). For
any $n$ Eq. (6) denotes a multi-well (-barrier) which includes the
well-known double-well with $n=1$. The absolute value $|g_{1d}|$
determines the required laser intensity. Obviously, Eq. (5)
describes the oscillating single soliton for $n=0$ and the soliton
trains for $n>0$ \cite{Strecker, Hai}. The oscillating amplitude
$x_0$ and frequency $\omega=1(\omega_x)$ of the solitonic center are
controlled by the oscillating laser strictly. The required laser
frequency $\omega_L$ is equal to the trapping frequency $\omega_x$,
that means resonance between the trapping and driving potentials.
The same frequency fixes width and height of the Gaussian laser and
solitons, through the axial length $l_x=\sqrt{\hbar /(m \omega_x)}$
of harmonic oscillator. On the other hand, rewriting Eq. (5) as
$\psi_n=\phi_n (x,t)e^{-i E_n t/\hbar}$, we obtain the Floquet
quasienergy $E_n=(\frac 1 2 +\mu+n )$ adjusted by the laser
parameter $\mu$. The corresponding Floquet state obeys
$\phi_n(x,t+2\pi)=\phi_n(x,t)$.

It is interesting noting that the soliton solutions of Eq. (5)
possess the same form for the attractive and repulsive interatomic
interactions. However, for different interactions the solitons are
associated with different shapes of the required Gaussian lasers.
Taking $n=0$ as an example, Eq. (6) exhibits that the laser barriers
correspond to the attractive interaction with $g_{1d}<0$ and the
laser wells are associated with the repulsive one with $g_{1d}>0$.
When we consider $N=10^4$ $^7$Li atoms with mass $m$ being $7$ times
of the proton mass $m_p$ and take the experimental parameters
\cite{Strecker} $a_s=\pm 1.5$nm, $\omega_x=20$Hz, $\omega_r=800$Hz,
the harmonic oscillator lengths and the interaction intensity become
$l_x=\sqrt{40}l_r =21.22\mu$m and $g_{1d}=2 N \omega_r a_s/(\omega_x
l_x)=\pm 56.55$.
\begin{figure}[htp]
\center
\includegraphics[width=1.5in]{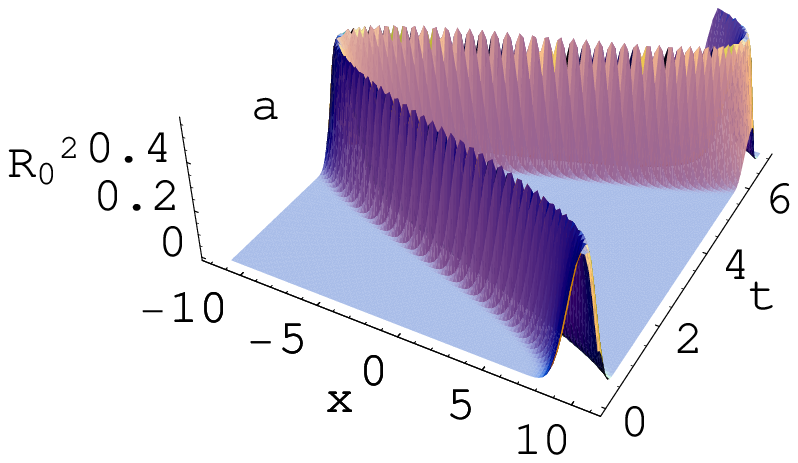}
\includegraphics[width=1.5in]{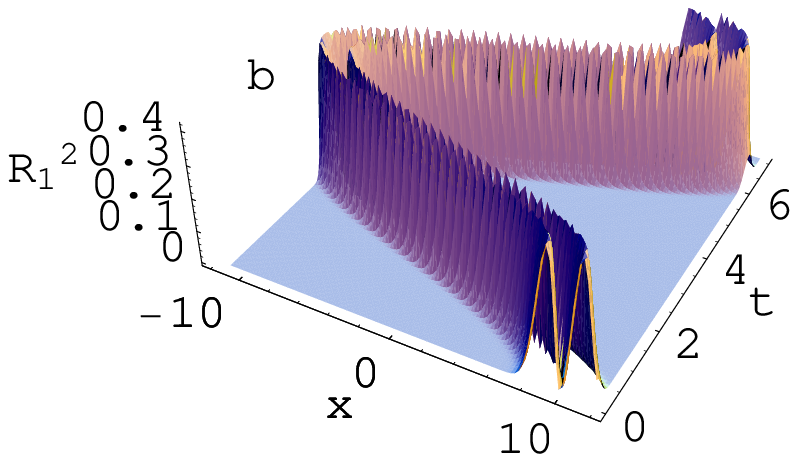}
\caption{Spatiotemporal evolutions of the atomic density plotted
from Eq. (5) for the laser parameter $x_0=10(l_x)=212.2\mu$m and (a)
$n=0$, (b) $n=1$. At any time Fig. 1a displays a single bright
soliton and Fig. 1b exhibits a soliton pair, both oscillating their
centers.}
\end{figure}
In Figs. 1a and 1b we show the spatiotemporal evolutions of the
atomic densities $R_0^2(x,t)$ and $R_1^2(x,t)$ respectively for the
laser parameter $x_0=10(l_x)=212.2\mu$m. The former is a single
Gaussian wave and the latter is a double Gaussian at any time for
both $g_{1d}>0$ and $g_{1d}<0$ cases. These bright solitons
oscillate their centers with amplitude $x_0$. Hereafter, all the
parameters in any figure are dimensionless.
\begin{figure}[htp]
\center
\includegraphics[width=1.5in]{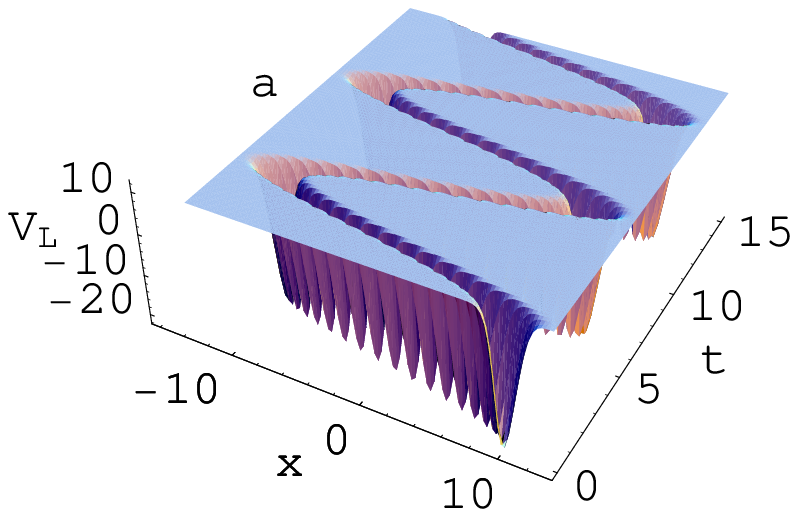}
\includegraphics[width=1.5in]{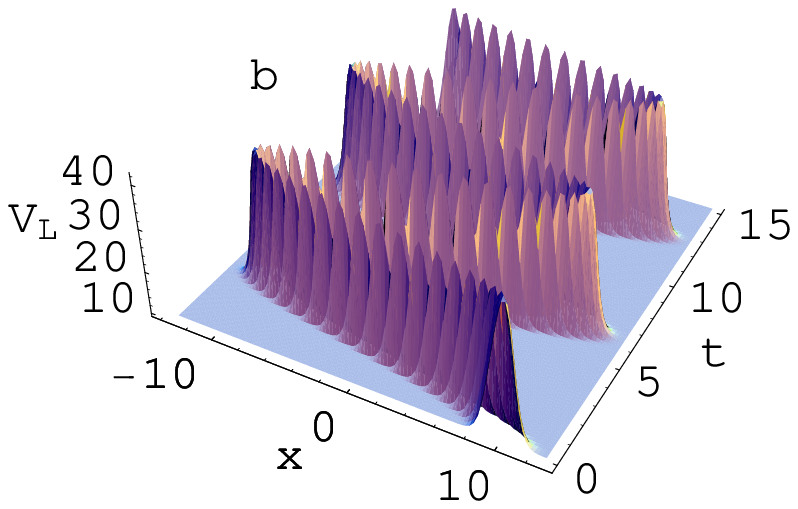}
\includegraphics[width=1.5in]{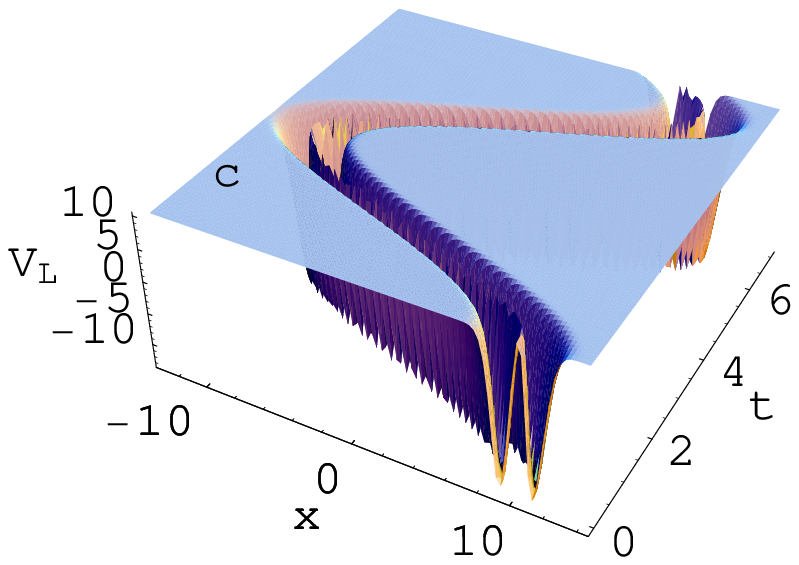}
\includegraphics[width=1.5in]{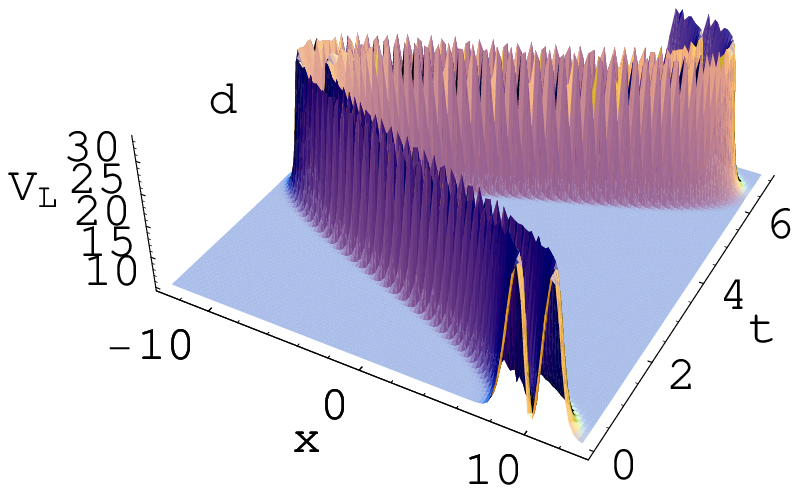}
\caption{Spatiotemporal evolutions of the potential functions
plotted from Eq. (6) for (a) $n=0,\ g_{1d}=56.55$, (b) $n=0,\
g_{1d}=-56.55$, (c) $n=1,\ g_{1d}=56.55$, (d) $n=1,\ g_{1d}=-56.55$.
The potential wells and barriers are found respectively for the
different cases.}
\end{figure}
The laser potential $V_L(x,t)$ of Eq. (6) with parameters $\mu=10,\
x_0=10$ is plotted as in Fig. 2 for four sets of parameters
respectively. In Figs. 2a and 2c with positive $g_{1d}$, we observe
the single and double wells respectively for any time. The wells
oscillate their center positions as the increase of time. From Figs.
2b and 2d with negative $g_{1d}$, we find that at any time the laser
potentials describe the single and double barriers with oscillating
centers.

\begin{figure}[htp]
\center
\includegraphics[width=1.4in]{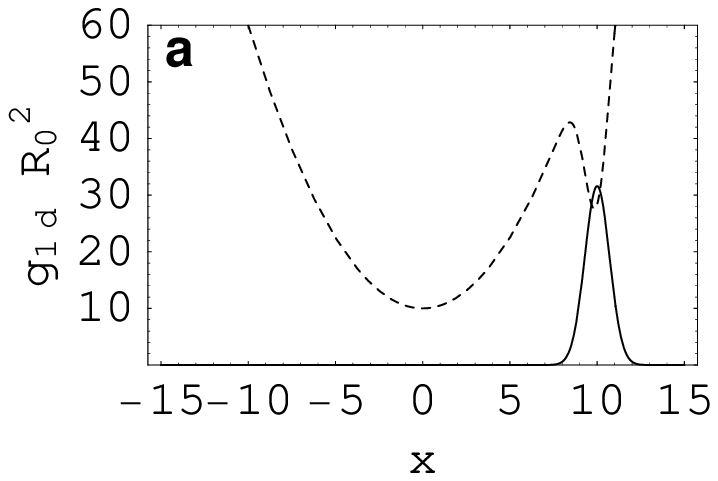}
\includegraphics[width=1.4in]{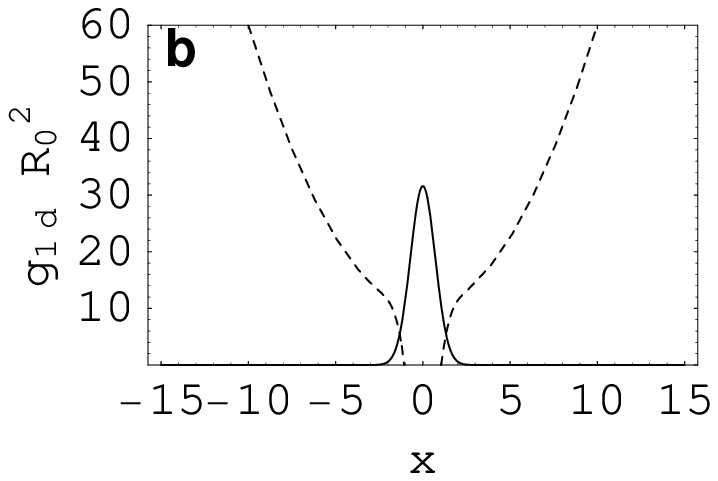}
\includegraphics[width=1.4in]{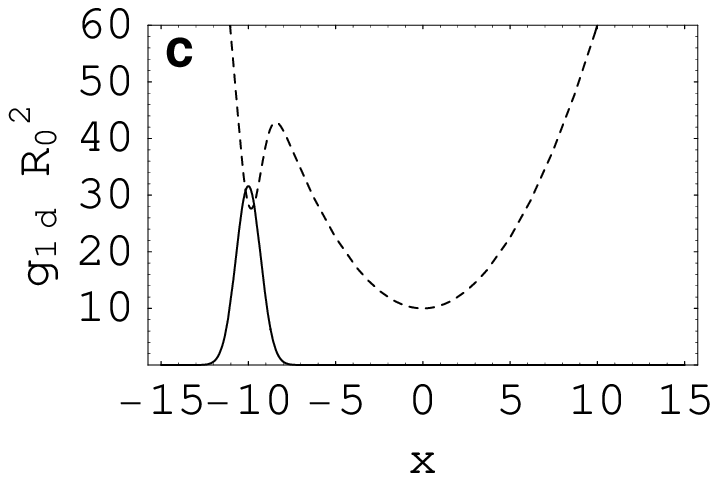}
\includegraphics[width=1.4in]{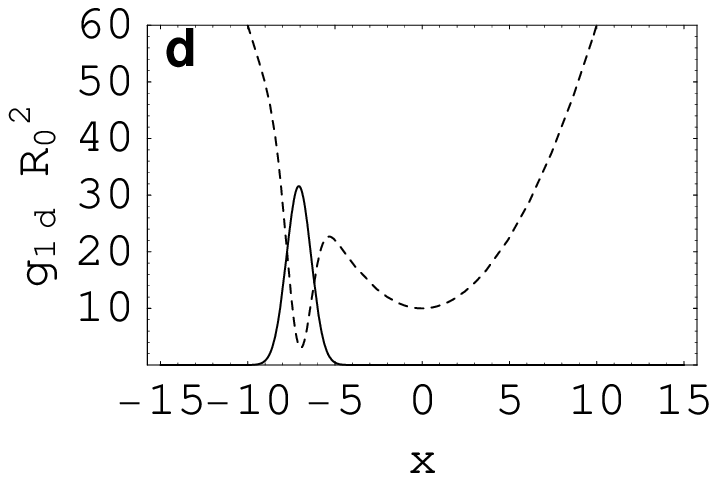}
\caption{Spatial profiles of the soliton (solid curves) and
potential functions (dashed curves) for the parameters of Fig. 2a
and at (a) $t=0$, (b) $t=\pi/2$, (c) $t=\pi$ and (d) $t=5\pi/4$. It
is observed that the soliton keeps its shape and oscillates its
position. The soliton center is located on the center of a potential
well for any time.}
\end{figure}
To see the details of the soliton motions and to analyze the
stability of the system, in Fig. 3 we show spatial profiles of the
soliton and total potential functions $V(x,t)=\frac 1 2
x^2+V_L(x,t)$ at several different times for the same parameters
with Fig. 2a. It is revealed that in the time evolution the
potential deforms between the single well and double well and the
density soliton keeps its initial shape. When the initial time is
taken as $t_0=0$, from Eq. (5) we plot the initial profile of
density times interaction intensity $g_{1d}R^2$ as in Fig. 3a, which
is a bright soliton centered at $x=10$. With time increasing to
$t=\pi/2$ and $t=\pi$, the soliton propagates toward left to
positions $x=0$ and $x=-10$ as in Figs. 3b and 3c. In such a time
interval, the potential deforms from a double well to a single well
then to another double well. In the next deformation period $\pi \le
\omega_x t \le 2\pi$ of the potential, the soliton propagates toward
right as shown in Fig. 3d and will back to the initial place. The
propagation property is in good agreement with that of Strecker's
experiment and the potential in Fig. 3a has the profile illustrated
in previous work \cite{Diener}. The soliton center is always located
on the center of a potential well. This implies that at any time as
a quasi-particle the soliton falls into a potential well. According
to the well-known criterion of dynamical stability, such soliton
motion is stable. In the transportation process of BEC soliton, the
laser potential $V_L(x,t)$ plays a role of optical tweezer
\cite{Diener,Gustavson}.

\begin{figure}[htp]
\center
\includegraphics[width=1.4in]{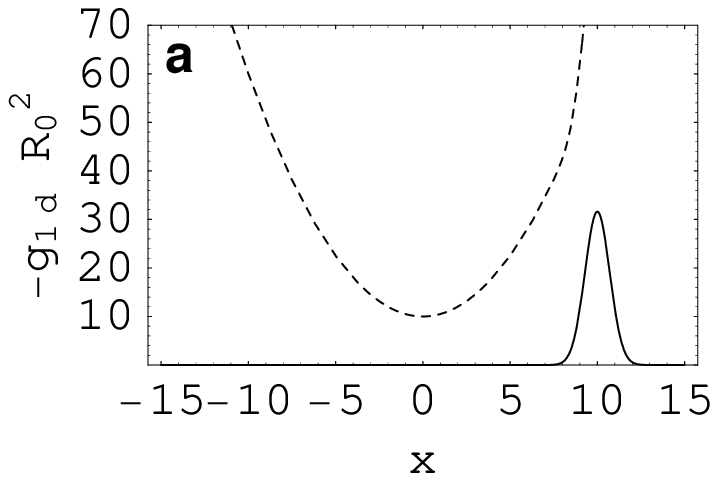}
\includegraphics[width=1.4in]{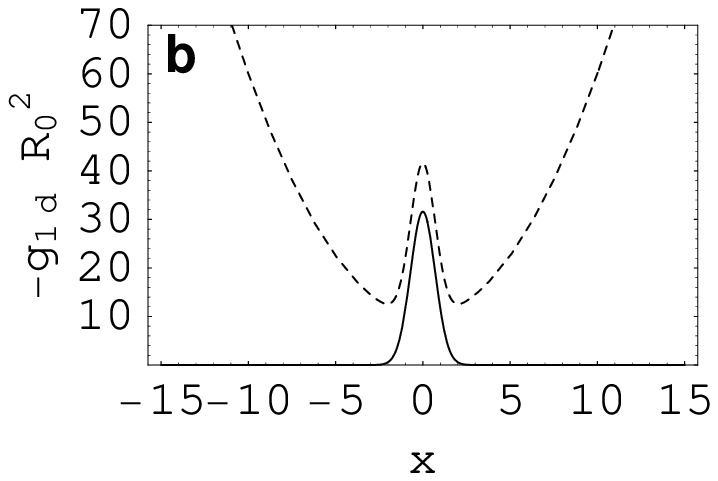}
\includegraphics[width=1.4in]{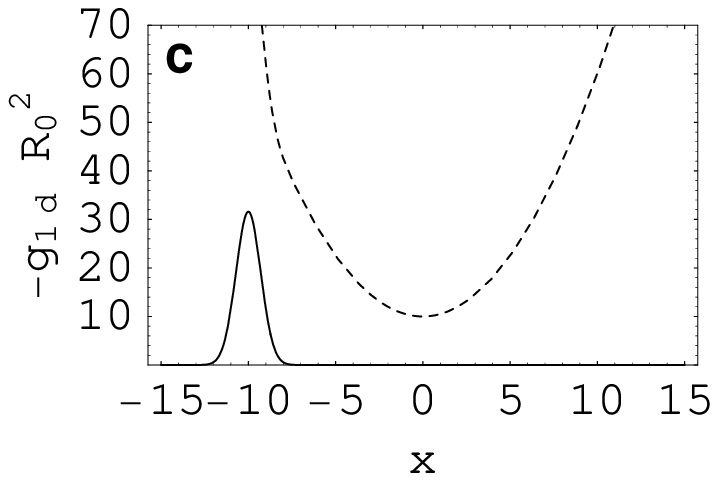}
\includegraphics[width=1.4in]{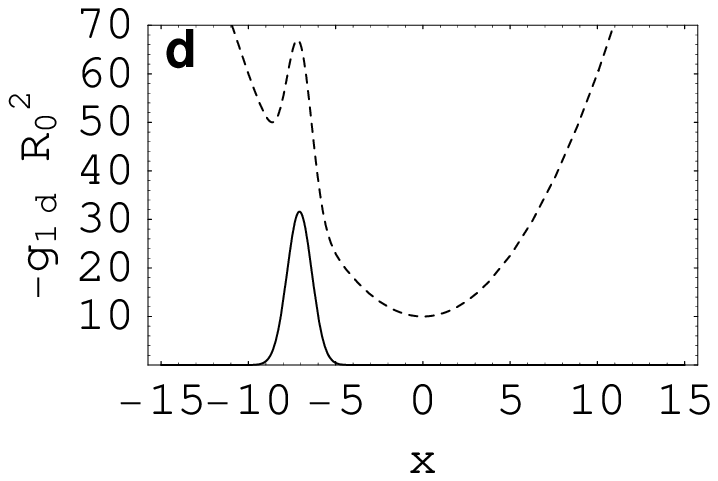}
\caption{Spatial profiles of the soliton (solid curves) and
potential functions (dashed curves) for the same parameters with
Fig. 2b and at (a) $t=0$, (b) $t=\pi/2$, (c) $t=\pi$ and (d)
$t=5\pi/4$. At $t=0$ and $t=\pi$ the soliton is not located on the
position of minimal potential. For $t=\pi/2$ and $t=5\pi/4$ the
soliton is located on the position of potential barrier.}
\end{figure}

When the interaction intensity is changed from $g_{1d}=56.55$ to
$g_{1d}=-56.55$, Fig. 3 is correspondingly changed to Fig. 4. In the
latter figure, the solitonic shape and evolution have no change.
However, the potential deforms with different pattern compared to
Fig. 3 such that the soliton is no longer located on the center of a
potential well. Particularly, Figs. 4b and 4d exhibit that sometimes
as a quasi-particle the soliton lies at the top of potential
barrier. The corresponding soliton motion may be dynamically
unstable thereby.

Similarly, in the double-soliton case with $n=1$, by comparing Fig.
1b with Figs. 2c and 2d, we find that the solitons fall on the
optical wells or barriers for the repulsive or attractive BEC.
Therefore, the exact soliton pair is dynamically stable for the
repulsive BEC or unstable for the attractive one.

In conclusion, we have investigated the repulsive and attractive
quasi-1D BECs held in the combination potential of the magnetic trap
and the Gaussian laser sheet with oscillating position. It is
demonstrated that when the laser potential balances the interatomic
interaction, the exact bright soliton trains can be generated. The
corresponding $n$ soliton solutions agree with the extended coherent
states of harmonic oscillator. The soliton trains fit the periodical
motions of laser centers and keep their shapes of Gaussian waves,
that agree with Strecker's experiment partly \cite{Strecker}. The
solitonic width, height, oscillating amplitude and frequency are
controlled by the laser field accurately. The required optical
potentials contain the Gaussian wells and barriers for the repulsive
and attractive BECs respectively, which resonate with the trapping
potential. For $n=0$ case the optical well is similar to the quantum
dot generated by a focused beam of red-detuned laser light
\cite{Diener}. Although the solitonic profile of fixed $n$ is same
for the both interaction cases, the soliton of repulsive BEC is more
stable than that of the attractive BEC. It is worth noting that in
some time intervals the BEC solitons are transported toward fixed
direction and the laser field behaves like a quantum tweezer
realized in previous works \cite{Diener,Gustavson}. Therefore, the
exact bright solitons could be observed and controlled
experimentally by oscillating the laser position and adjusting the
system parameters.

It is also noted that the exact soliton trains are similar to the
results of Ref. \cite{Beitia}. However, the required external
potentials and interatomic interactions are different for the both
cases. The spatiotemporal-dependent interaction intensity used in
\cite{Beitia} is not required in our systems. This could bring
convenience to the experimental observation of the soliton trains.

\begin{acknowledgments}
This work was supported by the National Natural Science Foundation
of China under Grant Nos. 10575034 and 10875039.
\end{acknowledgments}

\end{document}